\documentclass[aps,pra,twocolumn,superscriptaddress,showpacs,showkeys,amsmath,amssymb]{revtex4}

\usepackage{amsfonts}
\usepackage{amssymb,amsmath}
\usepackage{mathrsfs}
\usepackage{latexsym}
\usepackage{amsmath}
\usepackage[cp1251]{inputenc}
\usepackage{graphicx}
\usepackage{dcolumn}
\usepackage{bm}
\usepackage{color}

\begin{document}

    \title{Impurity self-energy in the strongly-correlated Bose systems}

    \author{G.~Panochko}
    \affiliation{College of Natural Sciences, 
    Ivan Franko National University of Lviv,\\ 107 Tarnavskyj  Str. Lviv, Ukraine\\}
    \author{V.~Pastukhov\footnote{e-mail: volodyapastukhov@gmail.com}}
    \author{I.~Vakarchuk}
    \affiliation{Department for Theoretical Physics, Ivan Franko National University of Lviv,\\
     12 Drahomanov Street, Lviv-5, 79005, Ukraine}

    \date{\today}



    \begin{abstract}
    We proposed the non-perturbative scheme for calculation of the impurity spectrum in the Bose system at zero temperature. The method is based on the path-integral formulation and describes an impurity as a zero-density ideal Fermi gas interacting with Bose system for which the action is written in terms of density fluctuations. On the example of the $^3$He atom immersed in the liquid helium-4 a good consistency with experimental data and results of Monte Carlo simulations is shown.
    \end{abstract}

    \maketitle

\section{Introduction}
\label{sec1}
\setcounter{equation}{0}

The concept of Bose polaron, i.e., a single impurity particle
immersed in the bosonic medium was originally introduced by Landau
and Pomeranchuk \cite{Landau_Pomeranchuk} in the context of the $^3$He
atom moving in the superfluid helium. In work \cite{Landau_Pomeranchuk} the
general structure of low-energy impurity spectrum was established
and shown to be dependent on two parameters, namely, a binding
energy and the atom effective mass. The latter is known \cite{Astrakharchik_Pitaevskii} to be uniquely
related to the depletion of the superfluid component in the Bose system due to interaction with impurity. Similarly to the original polaron problem the interaction-induced localization is also a characteristic of the impurity in homogeneous bosonic environment \cite{Cucchietti,Bruderer}. Recent renewal of an interest to the Bose polaron problem is associated with growing opportunities of experimental techniques \cite{Schmid_et_al,Spethmann_et_al}. Particularly in Ref.~\cite{Hu} (see also \cite{Jorgenzen}) the strongly-coupled impurity of $^{40}$K in the $^{87}$Rb Bose condensate was achieved. Besides their profound effect on the future study of mobile impurities in a bosonic medium and strongly interacting Bose systems these experimental results together with Monte Carlo simulations \cite{Ardila,Vlietinck} can serve as a test for numerous theoretical predictions \cite{Grusdt_Demler}, namely, variational \cite{Campbell,Tempere_et_al,Novikov_Ovchinnikov_10,Li} diagrammatic \cite{Novikov_Ovchinnikov,Rath,Christensen,Tkach,Loft} and recently published renormalization group \cite{Grusdt_et_al,Grusdt} and Hartree-Fock-Bogoliubov \cite{Kain_Ling} approaches. 

In the present paper by means of the self-consistent Green's function approach we have considered the impurity states in liquid $^4$He.

\section{Model and method description}
Instead of studying the properties of a single impurity atom
immersed in the non-ideal Bose system we consider a model of very
dilute spin-polarized ideal Fermi gas interacting with the Bose
condensed medium. This trick provides a possibility to use the
advantages of the field-theoretical approaches in order to solve
this problem. Adopting imaginary time path-integral formulation we
write down the action of our system, which principally contains
three terms
\begin{eqnarray}\label{S}
	S=S_0+S_B+S_{int}.
\end{eqnarray}
The first one
\begin{eqnarray}\label{S_0}
	S_0=\sum_{P}\{i\nu_p-\varepsilon_f(p)+\mu\}\psi^*_P\psi_P,
\end{eqnarray}
describes Fermi gas with dispersion $\varepsilon_f(p)=\hbar^{2}
p^2/2m_f$ and chemical potential $\mu$. 
Here $\psi^*_P$, $\psi_P$
are complex Grassmann fields, four-vector $P=(\nu_p,\textbf {p})$
where $\nu_p$ is the fermionic Matsubara frequency. The second
term in Eq.~(\ref{S}) is the action of the interacting Bose
particles. In order to simplify further consideration we adopt
Popov's density-phase formulation \cite{Popov} with integrated out
phase fields (see Ref.~\cite{Pastukhov_15} for details). The
appropriate action written in terms of density fluctuations
$\rho_K$ reads
\begin{eqnarray}\label{S_B}
	&&S_{B}=-\frac{1}{2}\sum_{K}D_0(K)\rho_K \rho_{-K}\\
	&&-\frac{1}{3!\sqrt{\beta V}}\sum_{K_1+K_2+ K_3=0}D_0(K_1, K_2,
	K_3)\rho_{K_1}\rho_{K_2}\rho_{K_3}\nonumber
\end{eqnarray}
where $K = (\omega_k,{\mathbf{
		k}})$ (note that ${\mathbf{k}}=0$ is
omitted in every summation over the wave-vector), $\omega_k$ is the
bosonic Matsubara frequency. 
We also introduced bare vertex
functions
\begin{eqnarray}\label{D_0(K)}
	D_0(K)=\frac{m \omega^2_k}{\rho \hbar^2 k^2}+\frac{\hbar^2k^2}{4m
		\rho}+\nu(k),
\end{eqnarray}
\begin{eqnarray}\label{D_0_3}
	D_0(K_1, K_2,
	K_3)=\frac{\hbar^2\mathbf{{
				k}}_1{\mathbf{
				k}}_2}{4m\rho^2}\left(1-\frac{\omega_{k_1}\omega_{k_2}}{\varepsilon_{k_1}\varepsilon_{k_2}}\right)+\textrm{perm.}
\end{eqnarray}
Microscopic parameters characterizing Bose subsystem $m$, $\rho$
and $\nu(k)$ are the mass of particles, an equilibrium density and the
Fourier transform of the two-body interaction, respectively. For
further convenience notation $\varepsilon_k=\hbar^2 k^2/2m$ is
introduced. We impose periodic boundary conditions with volume $V$
in a coordinate space and $\beta=1/T$ denotes the inverse
temperature of the system. The Gaussian term in Eq.~(\ref{S_B})
represents the action of non-interacting Bogoliubov
quasiparticles and the second term takes into account the simplest
collisional processes of these excitations.

Finally the last term of action (\ref{S}) that describes
interaction between fermions and Bose particles is
\begin{eqnarray}\label{S_int}
	S_{int}=-\rho \tilde{\nu}(0)\sum_P \psi^*_P\psi_P
	-\frac{1}{\sqrt{\beta V}}\sum_{K,P}\tilde{\nu}(k)\rho_K\psi^*_P\psi_{P-K},
\end{eqnarray}
where $\tilde{\nu}(k)$ is the Fourier transform of Bose-Fermi
two-body interaction and the first term can be absorbed by
shifting of the chemical potential $\tilde{\mu}=\mu-\rho
\tilde{\nu}(0)$. Exploring the behavior of an impurity spectrum
in the Bose condensate we have to consider fermionic single-particle
Green's function \cite{AGD}
\begin{eqnarray}
	G(P)=\langle \psi^*_P\psi_P\rangle=\left\{i\nu_p-\varepsilon_f(p)+\tilde{\mu}-\Sigma(P)\right\}^{-1},
\end{eqnarray}
(here $\langle \ldots\rangle$ denotes statistical averaging with
action (\ref{S})) where the {\it exact} self-energy $\Sigma(P)$ is given by a
single skeleton diagram depicted in Fig.~1.
\begin{figure}[h!]
	\centerline{\includegraphics
		[width=0.2\textwidth,clip,angle=-0]{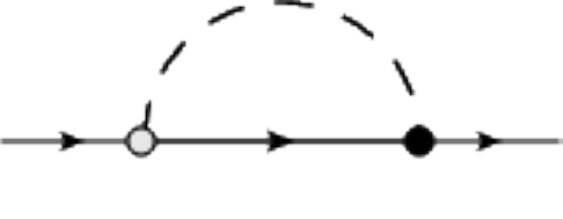}} \caption{Diagrammatic representation of the self-energy $\Sigma(P)$. The
		exact one-particle Green's function is denoted by bold solid line
		with arrow. For the exact bosonic density-density correlator the
		dashed line is used. Dots stand for zero-order (light) and exact
		(black) vertices, respectively.}
\end{figure}
This is formally exact equation that determines the impurity
Green's function. The perturbation theory is build for
boson-fermion vertex $\Gamma(P,P-K)$ which also functionally
depends on $G(P)$ and $\langle \rho_K \rho_{-K}\rangle$. As a
result, such a consideration leads to the system of coupled
non-linear integral equations which should be solved
self-consistently. Then, after analytical continuation in the
upper complex half-plane $\Sigma(P)_{i\nu_p \rightarrow
	\nu+i0}=\Sigma_R(\nu,p)+i\Sigma_I(\nu,p)$ one obtains the energy
$\varepsilon^*_f(p)=\varepsilon_f(p)+\Sigma_R(\varepsilon^*_f(p),p)-\Sigma_R(0,0)$
of the impurity with momentum ${\mathbf{
		p}}$. Here we have taken into
account the disappearance of radius of the Fermi sphere in the
extremely dilute (one-particle) limit, i.e.,
$\tilde{\mu}-\Sigma_R(0,0)\propto 1/V^{2/3}$. The imaginary part
$\Sigma_I(\nu,p)$ of the self-energy determines the damping of
impurity spectrum due to collisions with Bose particles. It is not
difficult to obtain a diagrammatic representation for
$\Sigma_I(\nu,p)$ by applying of the unitarity conditions
\cite{Maleev} for the diagram in Fig.~1. The resulting equation
can be written in terms of spectral weights of impurity and
phonons as well as the exact vertex $\Gamma(P,P-K)$ for which we
can use the exact estimation $\Gamma(P,P)=\tilde{\nu}(0)+\partial
\Sigma(P)/\partial \rho$ (the derivation is similar to that of
Ref.~\cite{Pastukhov_InfraredStr}). From the general arguments it
is clear that the creation of Bogoliubov excitations in the dilute
Bose gas at very low temperatures is only possible when the
impurity is moving through the superfluid with velocity $\hbar
p/m_f$ that is larger than the velocity $c$ of sound propagation.
In the liquid helium-4, however, due to presence of a roton minimum
in the excitation spectrum this momentum threshold is restricted
to a much smaller value \cite{Saarela_et_al}.

\section{Self-energy calculations}
The calculation scheme discussed previously is very hard for the
practical realization, therefore some approximation procedure
should be applied. We will assume that the information about the
exact density-density correlation function of Bose system is known
and treat only the fermionic self-energy perturbatively. In this
way by choosing the appropriate form of $\langle
\rho_K\rho_{-K}\rangle$ we are in position to predict the behavior
of impurity immersed both in the strongly-correlated Bose system
like liquid $^4$He as well as in the weakly-interacting Bose
condensates of alkali atoms.
\subsection{First-order results}
On the one-loop level the self-energy calculation is relatively
simple. Neglecting vertex corrections, i.e.
$\Gamma(P,P-K)=\tilde{\nu}(k)$, substituting simple ansatz for the
exact Green's function $G(P)=[i\nu_p-\varepsilon^*_f(p)]^{-1}$ and
making use of the Feynman approximation for the density-density
correlation function $\langle \rho_K
\rho_{-K}\rangle=2\rho\varepsilon_k/[\omega^2_k+E^2_k]$ (here
$E_k=\varepsilon_k\alpha_k$, where $1/\alpha_k=S_k$ is the static
structure factor of the Bose subsystem) we obtain in the
low-temperature limit
\begin{eqnarray}\label{Sigma_1}
	\Sigma^{(1)}(P)=-\frac{1}{V}\sum_{\mathbf{
			k}}\rho\frac{\tilde{\nu}^2(k)}{\alpha_k}\frac{1}{E_k+\varepsilon^*_f({\mathbf{
				k}}+{\mathbf{
				p}})-i\nu_p}.
\end{eqnarray}
We adopt the minimal substitution scheme with
$\varepsilon^*_f(p)=\hbar^2p^2/2m^*_f$, where the effective mass
$m^*_f$ should be calculated self-consistently. Such a choice of
the impurity Green's function simplifies further consideration
providing that all observables depend on one parameter $m^*_f/m$.
It is also believed that this substitution is well-grounded when
the major contribution to the integral in Eq.~(\ref{Sigma_1}) is
coming from the phonon region, i.e., when $\tilde{\nu}(k)$ is the
rapidly decreasing function of wave-vector. On the other hand, by
choosing the skeleton graph expansion even in the simplest
effective mass approximation we avoid the unphysical situation
with appearance of the finite magnitude damping \cite{Rath,Li,Panochko}
for the motionless particle. Note, that we also neglect the
renormalization of the quasiparticle residue $Z^{-1}(p)=1-\partial
\Sigma_R(\varepsilon^*_f(p),p)/\partial\varepsilon^*_f(p)$. The
presence of $Z(p)$ in our ansatz for the impurity Green's function
(where there is no imaginary part in the self-energy) immediately
brakes the particle number conservation law and enormously shifts the
integral characteristics of the impurity atom, like energy.

\subsection{Second-order results}
The second-order diagrams for the self-energy are depicted in Fig.~2.
\begin{figure}
	\centerline{\includegraphics
		[width=0.46
		\textwidth,clip,angle=-0]{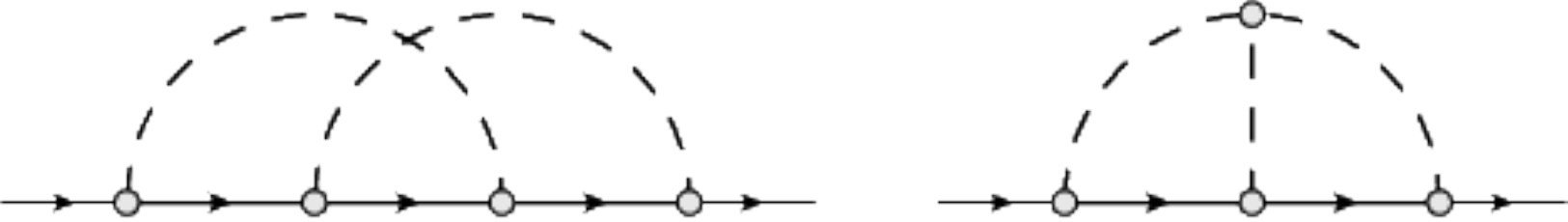}}
	\caption{Second-order diagrams contributing to the impurity self-energy $\Sigma(P)$.
	}
\end{figure}
They all stem from the boson-fermion vertex renormalization. We
recall that in our field-theoretic skeleton graph expansion we
have not to take into account the diagrams with the self-energy
insertions. It is also understood that for the full
self-consistency of the results one needs to solve the integral
equation for $\Gamma(P,P-K)$ in the appropriative approximation
and then substitute it in the formula for the self-energy. In the
language of Feynman's diagrams it particularly means that all
light dots which represent zero-order vertices with two fermionic
and one bosonic lines should be painted in black, i.e., replaced
by the exact one. But this program is very difficult for the
practical realization. Moreover, in this case we would also have
to substitute the exact three-legged bosonic vertex $D(K,Q,S)$ in
the impurity self-energy. The only information known about this
vertex is the infrared asymptote
$D(0,0,0)=\frac{\partial}{\partial \rho}\frac{mc^2}{\rho}$ and the
ultraviolet behavior where it coincides with (\ref{D_0_3}), which
is not enough for the proper evaluation of $\Sigma^{(2)}(P)$.
Therefore, in these second-order calculations we continue to use
our approximation scheme, where the Green's function $G(P)$ is
determined by the effective mass only, and the impurity-bosonic
vertex is treated perturbatively.  The explicit result of these straightforward but nevertheless cumbersome calculations of $\Sigma^{(2)}(P)$ at zero temperature is given in appendix.

\section{Numerical calculations and discussion}
In this work we considered the impurity states in the liquid
$^4$He. This is historically the first example of a
strongly-correlated Bose liquid for which the information about
the structure of excitations is well-measured \cite{Blagoveshchenskii}. The most
natural and experimentally relevant choice of the impurity atom is
$^3$He (or even $^6$He). Of course in the adopted approximations
the results of numerical calculations depend on the form of the
boson-impurity potential $\tilde{\nu}(k)$ only. In the adiabatic
approximation the two-body interaction of $^4$He-$^3$He is the
same as $^4$He-$^4$He, therefore we can easily express
$\tilde{\nu}(k)$ via the static structure factor of the superfluid
helium. In that way our calculations will not be restricted to the
isotopes of helium, but with some assumptions the obtained
formulae can be also used for the atoms of rare earth elements.
Thus, varying the mass $m_f$ we can study within our calculational
method the observable properties of diverse types of impurity
atoms immersed in the liquid $^4$He.

The simplest approximation for $\tilde{\nu}(k)$ is obtained through
the comparison of Bogoliubov's and Feynman's spectra
\begin{eqnarray}\label{nu_0}
	\tilde{\nu}(k)=\varepsilon_k (1/S^2_k-1)/2\rho.
\end{eqnarray}
Such a procedure of determination of the two-body potential allows \cite{all_together} to describe the properties of liquid helium-4 in the whole temperature region including second-order phase transition point.
Now by calculating $\Sigma_R(\varepsilon^*_f(p),p)-\Sigma_R(0,0)$
at small values of the wave-vector one derives the effective mass
Fig.~3, which was computed numerically at equilibrium
density $\rho=0.02185$ \AA$^{-3}$ of $^4$He. Then these curves were
used for calculations of the impurity atom binding energy
$\mu=\rho\tilde{\nu}(0)+\Sigma_R(0,0)$. The latter is
presented of the Fig.4.
\begin{figure}[h!]\label{fig.3}
	\includegraphics[scale=0.31]{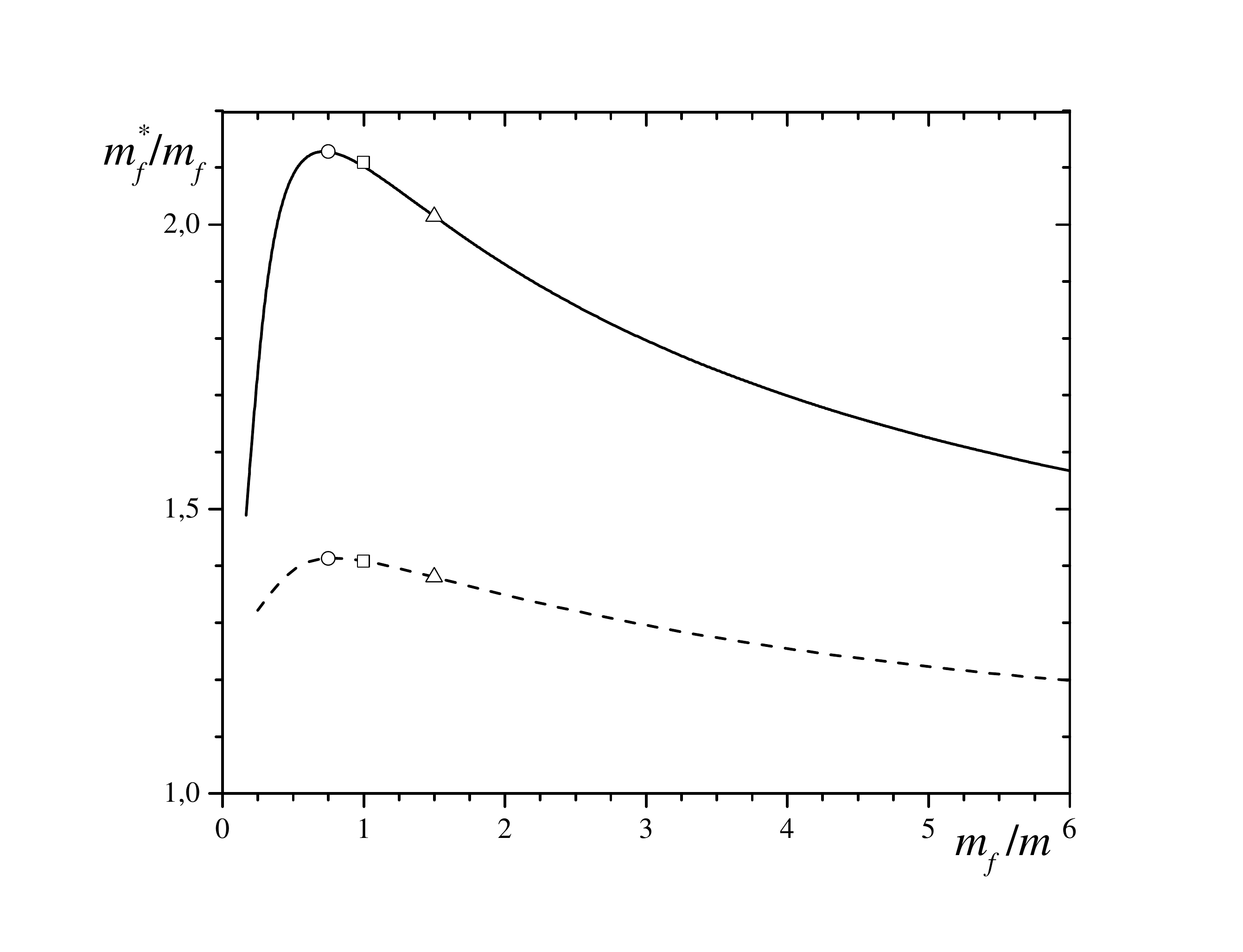}
	\caption{ Impurity effective mass in liquid helium-4 as a function of mass ratio $m_f/m$. In particular, circle -- $^3$He, square -- $^4$He, triange -- $^6$He impurities. The dashed and solid lines are the first- and second-order results, respectively.}
\end{figure}
\begin{figure}[h!]\label{fig.4}
	\includegraphics[scale=0.31]{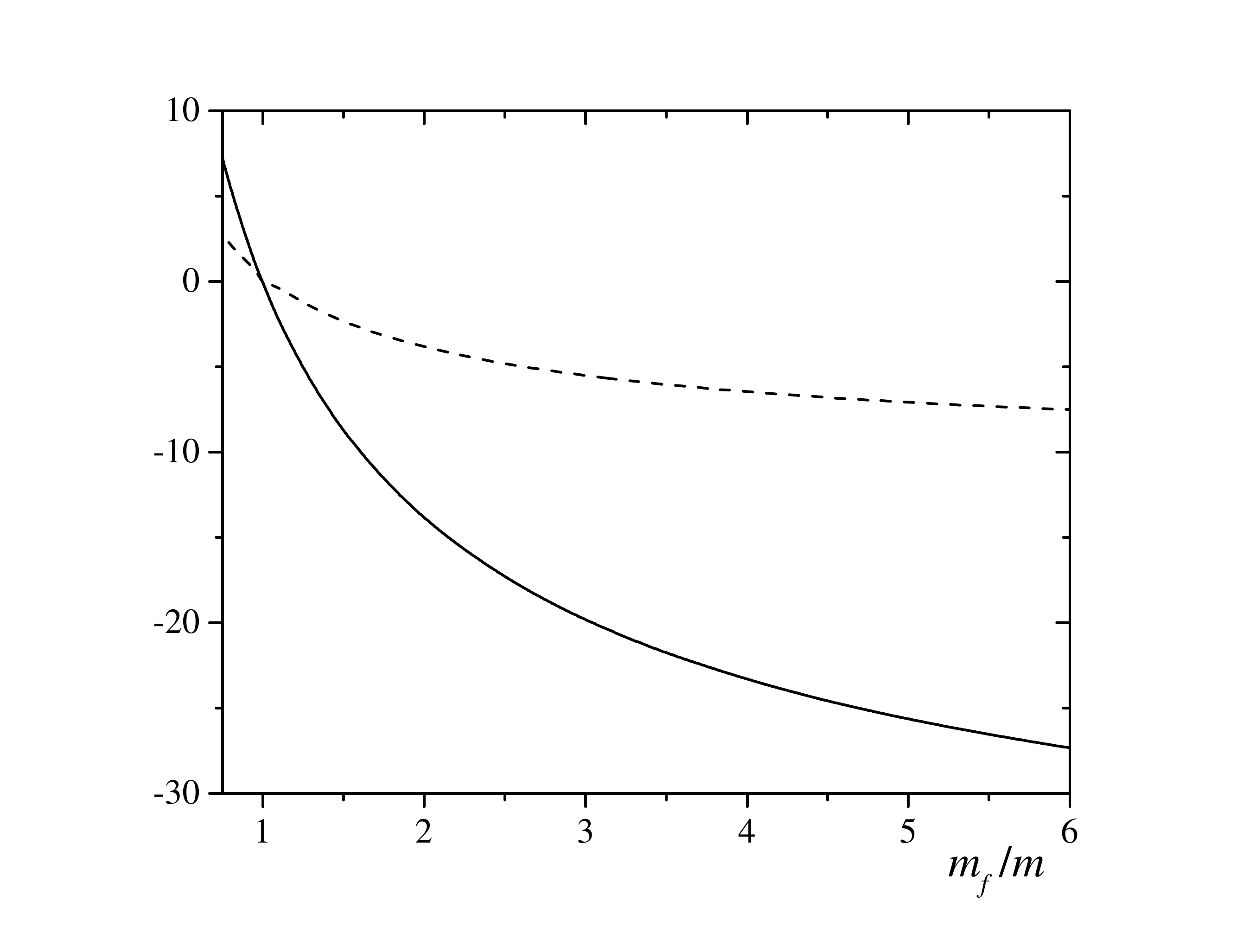}
	\caption{ Shifted energy $\mu(m_f/m)-\mu(1)$ (in Kelvins) of the impurity atom. The dashed and solid lines are the first- and second-order results, respectively.}
\end{figure}
The obtained within the first-order perturbation theory effective mass of $^4$He atom $m^*_f(1)/m_f=1.41$ can be compared to the result $1.58$ of\cite{Rovenchak}, where very similar self-consistent calculation scheme was used. Our second-order prediction for the effective mass of $^3$He $m^*_f(3/4)/m_f=2.13$ is in agreement with experimental values $2.18$ \cite{Yorozu}, $2.15$ \cite{Simons}, results of another theoretical studies $2.09$
\cite{Krotscheck_et_al}, diffusion  $2.20(5)$ \cite{Boronat} and variational $2.06$--$2.07$ \cite{Galli} Monte Carlo calculations.

\section{Conclusions}
In conclusion, by using path-integral formulation we have studied the properties of a single impurity atom immersed in the Bose system. We applied our approach which is based on skeleton-graph expansion to the problem of impurity states in the liquid helium-4. In particular, the effective mass as well as binding energy of an impurity atom with helium-like interparticle potential are calculated. The obtained value for the hydrodynamic mass of the $^3$He atom immersed in a Bose liquid coincides well both with experiments and results of Monte Carlo simulations. Finally, our calculations can be easily extended onto the low-dimensional systems.

\begin{center}
	{\bf Acknowledgements}
\end{center}
We thank Dr.~A.~Rovenchak and Dr.~O.~Hryhorchak for fruitful
discussions. This work was
partly supported by Project FF-30F (No.~0116U001539) from the
Ministry of Education and Science of Ukraine.

\section{Appendix}

In this section we present the explicit form of the second-order
impurity self-energy. Actually, the result of calculations of
two diagrams depicted in Fig.~2 is given. After simple integration
over the Matsubara frequencies with the use of a residue theorem
we obtain
\begin{widetext}
\begin{eqnarray}\label{Sigma_2}
\Sigma^{(2)}(P)=\frac{1}{V^2}\sum_{{\bf k},{\bf s}}\rho\frac{\tilde{\nu}(k)\tilde{\nu}(s)\tilde{\nu}(|{\bf k}+{\bf s}|)}{\alpha_k\alpha_s\alpha_{|{\bf k}+{\bf s}|}}\frac{1}{E_s+\varepsilon^*_f({\bf s}-{\bf p})-i\nu_p}\frac{1}{E_k+\varepsilon^*_f({\bf k}+{\bf p})-i\nu_p}\nonumber\\
-\frac{1}{V^2}\sum_{{\bf k},{\bf s}}\rho^2\frac{\tilde{\nu}^2(k)\tilde{\nu}^2(s)}{\alpha_k\alpha_s}\frac{1}{E_k+\varepsilon^*_f({\bf k}+{\bf p})-i\nu_p}\frac{1}{E_s+\varepsilon^*_f({\bf s}+{\bf p})-i\nu_p}\nonumber\\
\times \frac{1}{E_k+E_s+\varepsilon^*_f({\bf k}+{\bf s}+{\bf p})-i\nu_p}\nonumber\\
-\frac{1}{V^2}\sum_{{\bf k},{\bf s}}\rho\frac{\tilde{\nu}(k)\tilde{\nu}(s)\tilde{\nu}(|{\bf k}+{\bf s}|)}{\alpha_k\alpha_s\alpha_{|{\bf k}+{\bf s}|}}\frac{D_{+}({\bf k},{\bf s})}{E_{|{\bf k}+{\bf s}|}+\varepsilon^*_f({\bf k}+{\bf s}+{\bf p})-i\nu_p}\frac{1}{E_k+\varepsilon^*_f({\bf k}+{\bf p})-i\nu_p}\nonumber\\
\times\frac{1}{E_k+E_s+\varepsilon^*_f({\bf k}+{\bf s}+{\bf p})-i\nu_p}\nonumber\\
+\frac{1}{V^2}\sum_{{\bf k},{\bf s}}\rho\frac{\tilde{\nu}(k)\tilde{\nu}(s)\tilde{\nu}(|{\bf k}+{\bf s}|)}{\alpha_k\alpha_s\alpha_{|{\bf k}+{\bf s}|}}\frac{D_{-}({\bf k},{\bf s})}{E_k+E_s+E_{|{\bf k}+{\bf s}|}}\frac{1}{E_k+\varepsilon^*_f({\bf k}+{\bf p})-i\nu_p}\nonumber\\
\times\frac{1}{E_k+E_s+\varepsilon^*_f({\bf k}+{\bf s}+{\bf p})-i\nu_p},
\end{eqnarray}
where the symmetric functions $D_{\pm}({\bf k},{\bf s})$ read
\begin{eqnarray*}
	D_{\pm}({\bf k},{\bf s})=\frac{\hbar^2}{2m}\left[{\bf k}({\bf k}+{\bf s})(\alpha_k-1)(\alpha_{|{\bf k}+{\bf s}|}\pm1)+{\bf s}({\bf s}+{\bf k})(\alpha_s-1)(\alpha_{|{\bf k}+{\bf s}|}\pm1)\pm{\bf k}{\bf s}(\alpha_k-1)(\alpha_s-1)\right].
\end{eqnarray*}
\end{widetext}

\end{document}